\documentclass[12pt]{article}
\usepackage{jheppub}
\usepackage{amssymb}
\usepackage{mathrsfs} %script fonts
\usepackage{graphicx}
\usepackage{bm}% bold math
\usepackage{wasysym}
\def\A{{\mathscr{A}}}

\title{Entropy bounds in terms of the $w$ parameter}

\author{Gabriel Abreu$^1$, Carlos Barcel\'o$^2$, {\rm and} Matt Visser$^1$}

\affiliation{$^1$ School of Mathematics, Statistics, and Operations Research,\\
Victoria University of Wellington, PO Box 600, Wellington 6140, New Zealand}
\affiliation{$^2$ Instituto de Astrof\'{\i}sica de Andaluc\'{\i}a, IAA--CSIC, Glorieta de la Astronom\'i{}a,\\
18008 Granada, Spain}

\emailAdd{gabriel.abreu@msor.vuw.ac.nz} 
\emailAdd{carlos@iaa.es}
\emailAdd{matt.visser@msor.vuw.ac.nz}

\abstract{
In a pair of recent articles [PRL {\bf 105} (2010) 041302; JHEP {\bf 1103} (2011) 056] two of the current authors have developed an entropy bound for equilibrium uncollapsed matter using only classical general relativity, basic thermodynamics, and the Unruh effect. An odd feature of that bound, $S\leq {1\over2} \A$, was that the proportionality constant, ${1\over2}$, was weaker than that expected from black hole thermodynamics, ${1\over4}$.  In the current article we strengthen the previous results by obtaining a bound involving the (suitably averaged) $w$~parameter. 
Simple causality arguments restrict this averaged $\langle w \rangle$ parameter to be $\leq 1$. When equality holds, the entropy bound saturates at the value expected based on black hole thermodynamics.
We also add some clarifying comments regarding the (net) positivity of the chemical potential. Overall, we find that even in the absence of any black hole region, we can nevertheless get arbitrarily close to the Bekenstein entropy.
  }
\keywords{
Entropy bounds, uncollapsed matter, $w$ parameter.\\
13 September 2011;  23 September 2011; 26 November 2011;\\
 \LaTeX-ed \today}

\begin{document}
%-------------------------------------------------------------------------
\maketitle
\flushbottom

%----------------------------------------------------
\def\implies{\Rightarrow}
%----------------------------------------------------
\renewcommand{\sun}{\ensuremath{\odot}}%
\def\ie{{\emph{i.e.}}}
\def\Barcelo{Barcel\'o}
\def\d{{\mathrm{d}}}
\def\etc{\emph{etc}}
\def\A{{\mathscr{A}}}
%----------------------------------------------------
\section{Introduction}
%----------------------------------------------------

In a pair of recent articles~\cite{Abreu1, Abreu2}, two of the current authors demonstrated that a blob of uncollapsed matter in internal equilibrium and with surface area $\A$ always satisfies
\begin{equation}
S \leq {1\over2} \; k_B \; {\A\over L_P^2} = {k_B\; \A\over 2 \, G_N \, \hbar},
\end{equation}
which in usual theoretician's units becomes
\begin{equation}
S \leq {\A\over 2}.
\end{equation}
The derivation of that bound depended only on classical general relativity (for uncollapsed sources, no black holes are involved in the argument), some basic thermodynamics, and the Unruh effect. (With the Unruh effect being the only step at which semiclassical quantum physics entered.) See also references~\cite{Abreu3, Abreu4, Abreu5} for a more detailed discussion. (From a somewhat different perspective, it is also interesting to realize~\cite{Kolekar:2010} that the entropy of a box of gas in fixed black hole geometry changes from a volume to an area law as the box is lowered toward the black-hole horizon.) 

An annoying feature of the bound derived in~\cite{Abreu1, Abreu2} was the presence of the numerical factor ${1\over2}$, since black hole thermodynamics~\cite{mechanix, Bekenstein-gsl,Hawking} would seem to  suggest that ${1\over4}$ might be a more appropriate numerical factor. Indeed the derivations presented in~\cite{Abreu1, Abreu2} used a number of sub-optimal inequalities, and in this article we shall tighten the bound considerably, with the improved bound being characterized by a (suitably averaged version of) the $w$ parameter. 

For purely pedagogical reasons and clarity of presentation it is convenient to first treat $w$ as a constant, specifying a linear  equation of state, $p=w\; \rho$. Subsequently, we treat $w$ as simply a ``descriptor'', $w(\rho)=p(\rho)/\rho$, which can now depend on the unspecified equation of state in some complicated manner. 
We shall demonstrate that (after suitable averaging)
\begin{equation}
S \leq {1+\langle w \rangle\over 1+3\langle w \rangle}  \; {\mathscr{A}\over 2}.
\end{equation}
For diffuse systems (where $\langle w \rangle \gtrsim 0 $) this reproduces the bounds of references~\cite{Abreu1, Abreu2, Abreu3}, and~\cite{Abreu5}. 
Simple causality arguments (see appendix~\ref{appendix:w}) imply that we must always have $w \leq 1$, and thus $\langle w \rangle \leq 1$.
Then, for ``stiff'' systems (where $\langle w \rangle \lesssim 1$) the bound becomes arbitrarily close to the black hole thermodynamic bound $S={1\over4}\A$, see for instance references~\cite{mechanix, Bekenstein-gsl, Hawking}.

For purely technical reasons static (nonrotating) and stationary (rotating) cases need to be dealt with separately, though consideration of the static case is sufficient to capture the key ideas. We also add some clarifying comments regarding the (net) positivity of the chemical potential, specifically in the case where the uncollapsed matter contains a particle-antiparticle mixture (see appendix~\ref{appendix:chemical}). This is a technical point required at one stage of the proof.

In summary, we now have an improved bound, and an improved proof, that gets arbitrarily close to the black hole bound.

%------------------------------------------------------------------------------------------------------------------------------------------
\section{The $w$ parameter as equation of state}
%------------------------------------------------------------------------------------------------------------------------------------------

For the sake of discussion, let us begin by imposing a very special linear equation of state  $p = w \rho$.
(This is an extremely restrictive assumption, but it is pedagogically transparent, and we shall  improve on this argument later in the article.)

%------------------------------------------------------------------------------------------------------------------------------------------
\subsection{Static (non-rotating) case}
%------------------------------------------------------------------------------------------------------------------------------------------

In a static geometry the Tolman mass is given by
\begin{equation}
m_T =  \int_\Omega \sqrt{-g_4} \{ \rho+3p\} \d^3 x.
\end{equation}
(See for instance reference~\cite{Abreu1},  equation (2).)
Furthermore by integrating the Gibbs--Duhem relation, while demanding Tolman--Ehrenfest equilibrium~\cite{TE, Tolman-35, Klein, Tolman-30, Tolman-book}, one has
\begin{equation}
S = {1\over T_\infty} \int_\Omega \sqrt{-g_4} \{ \rho+p\} \d^3 x - {\mu_\infty N\over T_\infty}.
\end{equation}
(See reference~\cite{Abreu1},  equation (19), and the accompanying discussion.)
Assuming the linear equation of state $p = w\; \rho$ we see
\begin{equation}
m_T =  \{1+3w\} \int_\Omega \sqrt{-g_4} \; \rho \; \d^3 x,
\end{equation}
while 
\begin{equation}
S = {1+w\over T_\infty} \int_\Omega \sqrt{-g_4} \; \rho\;  \d^3 x - {\mu_\infty N\over T_\infty}.
\end{equation}
Therefore
\begin{equation}
S = {1+w\over 1+3w } \; { m_T\over T_\infty} - {\mu_\infty N\over T_\infty}.
\end{equation}
Up to this stage we have only used internal thermodynamic equilibrium (via the Tolman--Ehrenfest relations), basic thermodynamics as encoded in the Gibbs--Duhem relation, and the definition of Tolman mass. 
Assuming positive chemical potential we obtain
\begin{equation}
S \leq {1+w\over 1+3w}  \; { m_T\over T_\infty}. 
\end{equation}
(Some technical comments on the (net) positivity of the chemical potential are relegated to appendix~\ref{appendix:chemical}.)
Now apply the theorem encoded in equation (13) of reference~\cite{Abreu1}, (this is where the Einstein equations enter). This theorem relates the Tolman mass to an integral of the surface gravity over the boundary of the uncollapsed body and in the present context implies
\begin{equation}
S \leq {1+w\over 1+3w}  \; { \int \kappa \; \d\mathscr{A}\over 4\pi \; T_\infty}. 
\end{equation}
Finally apply the temperature bound $T_\infty \geq {1\over2\pi} \max_{\mathscr A} \kappa$ encoded in equation (24) of reference~\cite{Abreu1}, (this is where semi-classical quantum physics enters).  Full technical details can be found in~\cite{Abreu1}.
We finally have
\begin{equation}
S \leq {1+w\over 1+3w}  \; {\mathscr{A}\over 2}.
\end{equation}
Now, note that for causally well behaved system we cannot have $w>1$; for uncollapsed matter we must (for a causal energy flux) have $w\in[-1,1]$ and (for a causal speed of sound) have $w\in[0,1]$. In either case we certainly  have $w\leq1$. A technical discussion of this point is relegated to appendix~\ref{appendix:w}. This condition is saturated by ``stiff matter'' ($w=1$). In this case we found precisely the same bound that arises in black hole thermodynamics
\begin{equation}
S \leq  {\mathscr{A}\over 4}.
\end{equation}
But as derived this bound requires a very special and very restrictive equation of state, so we shall subsequently relax the conditions under which a similar bound can be extracted. (For a detailed discussion of the star-like configurations one might expect to arise from this equation of state, see for instance section 4 of reference~\cite{Yunes}.) 

%------------------------------------------------------------------------------------------------------------------------------------------
\subsection{Stationary (rotating) case}
%------------------------------------------------------------------------------------------------------------------------------------------

Several technical steps in the discussion now change. 
From reference~\cite{Abreu1} we still have equation~(19)
\begin{equation}
S = {1\over T_\infty} \int_\Omega \sqrt{-g_4} \{ \rho+p\} \d^3 x - {\mu_\infty N\over T_\infty},
\end{equation}
which is now reference~\cite{Abreu2} equation (5.8).  The details of the derivation are somewhat different, but the result is the same.
We can therefore still assert
\begin{equation}
S = {1+w \over T_\infty} \int_\Omega \sqrt{-g_4} \; \rho \;\d^3 x - {\mu_\infty N\over T_\infty}.
\end{equation}
On the other hand, the Tolman mass is no longer given by reference~\cite{Abreu1} equation (3). Instead we have a different result that
\begin{equation}
\int_\Omega \sqrt{-g_4} \{ \rho+3p\} \d^3 x \leq {1\over4\pi} \int \{\kappa\cdot \hat n\} \sqrt{-g_2} \d^2x.
\end{equation}
See reference~\cite{Abreu2} equations (6.5) through (6.17). This allows us to assert
\begin{equation}
\{1+3w \} \int_\Omega \sqrt{-g_4} \rho \d^3 x \leq {1\over4\pi} \int \{\kappa\cdot \hat n\} \sqrt{-g_2} \d^2x
\leq {||\kappa||_\mathrm{max} \, \mathscr{A}\over 4 \pi},
\end{equation}
whence
\begin{equation}
S \leq  {1+w \over  1+ 3 w } \; {||\kappa||_\mathrm{max} \over 4 \pi  T_\infty }   \; \mathscr{A}
     - {\mu_\infty N\over T_\infty}.
\end{equation}
Again taking the chemical potential to be positive, see appendix~\ref{appendix:chemical}, and using our temperature bound~\cite{Abreu2}, we see
\begin{equation}
S \leq  {1+ w \over  1+ 3w} \; { \mathscr{A}\over 2}.
\end{equation}
Although the logic was somewhat different, this is the same result we just obtained for the static case.

%------------------------------------------------------------------------------------------------------------------------------------------
\section{The $w$ parameter as descriptor}
%------------------------------------------------------------------------------------------------------------------------------------------

Now define $w = p/\rho$; this is now not an equation of state, merely a definition of an interesting ratio to consider.
Furthermore, define a mass-weighted average
\begin{equation}
\langle w \rangle = {\int_\Omega \sqrt{-g_4} \; w \rho \; \d^3 x \over \int_\Omega \sqrt{-g_4} \; \rho\;  \d^3 x } =
{\int_\Omega \sqrt{-g_4} \; p\;  \d^3 x  \over \int_\Omega \sqrt{-g_4} \; \rho\;  \d^3 x }.
\end{equation}
We shall now see that using this averaged $\langle w \rangle$ yields an entropy bound very close in spirit to that above. 

%------------------------------------------------------------------------------------------------------------------------------------------
\subsection{Static (non-rotating) case}
%------------------------------------------------------------------------------------------------------------------------------------------

In terms of this average $\langle w \rangle$ the Tolman mass becomes
\begin{equation}
m_T =  \{1+3 \langle w \rangle\} \int_\Omega \sqrt{-g_4} \; \rho \; \d^3 x.
\end{equation}
(See for instance equation (2) of reference~\cite{Abreu1}.)
Similarly,  equation (19) of reference~\cite{Abreu1} becomes
\begin{equation}
S = {1+\langle w \rangle\over T_\infty} \int_\Omega \sqrt{-g_4} \; \rho \;\d^3 x - {\mu_\infty N\over T_\infty}.
\end{equation}
whence
\begin{equation}
S = {1+\langle w \rangle\over 1+3\langle w \rangle} \; { m_T\over T_\infty} - {\mu_\infty N\over T_\infty}.
\end{equation}
Apply the same logic as for the previous section, (positive chemical potential, the integral theorem for Tolman mass in terms of surface gravity, and the semiclassical temperature bound), we now have
\begin{equation}
S \leq {1+\langle w \rangle\over 1+3\langle w \rangle}  \; {\mathscr{A}\over 2}.
\end{equation}
This is just the previous result with $w \to \langle w \rangle$. 
Now in diffuse systems we know $\langle w \rangle\approx 0$ and are back to $S=\mathscr{A}/ 2$. However if the system is close to collapse, with most of the mass in the system being close to the upper limit of ``stiff matter'' ($w=1$) then $\langle w \rangle \lesssim 1$ and  it is useful to rewrite this as
\begin{equation}
S \leq \left\{ 1 + {1-\langle w \rangle\over 1+3\langle w \rangle} \right\} \; {\mathscr{A}\over 4}  \approx  {\mathscr{A}\over 4}. 
\end{equation}
This gives us our missing factor of 2 --- but only for ``close-to-collapse'' systems where most of the matter is close to the upper limit of $w\approx 1$. 
Note that 
\begin{equation}
{\d\over \d \langle w \rangle}\left({1+\langle w \rangle\over 1+3\langle w \rangle}\right) = - {2\over(1+3\langle w \rangle)^2} < 0.
\end{equation}
That is, the coefficient arising in this entropy bound is monotone decreasing,  and approaches ${1\over4}$ as as $\langle w \rangle$ increases to the stiff matter limit $\langle w \rangle=1$. 

%------------------------------------------------------------------------------------------------------------------------------------------
\subsection{Stationary (rotating) case}
%------------------------------------------------------------------------------------------------------------------------------------------

We still have reference~\cite{Abreu1} equation (19), reference~\cite{Abreu2} equation (5.8),
\begin{equation}
S = {1\over T_\infty} \int_\Omega \sqrt{-g_4} \{ \rho+p\} \d^3 x - {\mu_\infty N\over T_\infty}.
\end{equation}
So we can assert
\begin{equation}
S = {1+\langle w \rangle\over T_\infty} \int_\Omega \sqrt{-g_4} \; \rho \;\d^3 x - {\mu_\infty N\over T_\infty}.
\end{equation}
We still have (see reference~\cite{Abreu2} equations (6.5) through (6.17)), 
\begin{equation}
\int_\Omega \sqrt{-g_4} \{ \rho+3p\} \d^3 x \leq {1\over4\pi} \int \{\kappa\cdot \hat n\} \sqrt{-g_2} \d^2x.
\end{equation}
This allows us to assert
\begin{equation}
\{1+\langle 3w\rangle \} \int_\Omega \sqrt{-g_4} \rho \d^3 x \leq {1\over4\pi} \int \{\kappa\cdot \hat n\} \sqrt{-g_2} \d^2x 
\leq {||\kappa||_\mathrm{max} \mathscr{A}\over 4 \pi},
\end{equation}
whence
\begin{equation}
S \leq  {1+\langle w \rangle\over  1+ 3\langle w \rangle} \; {||\kappa||_\mathrm{max} \over 4 \pi  T_\infty }   \mathscr{A}  - {\mu_\infty N\over T_\infty}.
\end{equation}
Again taking the chemical potential to be positive, and using our temperature bound, we again have
\begin{equation}
S \leq  {1+\langle w \rangle\over  1+ 3\langle w \rangle} \; { \mathscr{A}\over 2}.
\end{equation}

%----------------------------------------------------
\section{Discussion}
%----------------------------------------------------

A particularly nice feature of the current analysis is that even for uncollapsed matter in internal equilibrium one can now get arbitrarily close to the bound expected on the basis of black hole thermodynamics, \emph{provided one allows the uncollapsed matter to approach the ``stiff matter'' limit} ($w\to1$). More precisely, since to approach the black hole entropy limit we require $\langle w \rangle \to 1$, and since $\langle w \rangle$ is a weighted average of $w$, we need $w\to1$ everywhere in the blob of matter under consideration. 

This has implications for any gravastar/ monster/ black-star/ quasi-black-hole one might wish to construct as a potential alternative to usual black holes~\cite{gravastar, gravastar2, monster, black-star, minimal, pseudo, qbh}. If you wish to use one of these ``black hole alternatives'' to develop a microscopic interpretation/ justification  for the usual black hole entropy, then one seems unavoidably driven towards the ``stiff matter'' limit --- throughout the blob of matter one is considering. In this regard the model of Lemos and Zaslavskii seems particularly interesting~\cite{qbh}.

It is also interesting to recall that there are ways in which black hole mimics, bodies without strict trapping horizons, could also evaporate by emitting Hawking-like radiation~\cite{barbado, barcelo-rad-not-trapped, thooft}. 
At the end of the day, it is not unreasonable to think that these uncollapsed objects might turn out to implement various thermodynamic properties equivalent to  standard black hole physics. In that case,  the black hole picture might  still be considered a useful effective model for analyzing at least some aspects of the behaviour of these uncollapsed black hole mimics.  

%-------------------------------------------------
\appendix
%------------------------------------------------------------------------------------------------------------------------------------------

%------------------------------------------------------------------------------------------------------------------------------------------
\section{Why is $w \leq 1$?}
\label{appendix:w}
%------------------------------------------------------------------------------------------------------------------------------------------

There are two main arguments commonly advanced for asserting $w\leq 1$:
\begin{itemize}
\item Energy conditions.
\item Speed of sound.
\end{itemize}
The energy condition arguments are weak and (without further modifications and suitable extensions) not to be trusted~\cite{Barcelo:twilight}. The speed of sound argument can (under suitable circumstances) be made rigorous. 

%------------------------------------------------------------------------------------------------------------------------------------------
\subsection{Energy condition argument}
%------------------------------------------------------------------------------------------------------------------------------------------

The usual classical energy conditions may not be fundamental physics~\cite{Barcelo:twilight}, but they give some quick and dirty (and useful) bounds on pressures and densities:
\begin{itemize}
\item The NEC implies $\rho+p\geq0$, which then implies $w\geq -1$ if $\rho>0$, \\ but $w\leq -1$ if $\rho<0$.
\item The WEC implies  $\rho+p\geq0$ and $\rho>0$, which then  implies $w\geq -1$.
\item The SEC implies $\rho+3p\geq0$ and  $\rho+p>0$, which then  implies  $w\geq -1/3$ if $\rho>0$, but $w\leq -1$ if $\rho<0$.
\item The DEC implies $\rho\geq0$ and $p\in[-\rho,+\rho]$, which then implies $w \in [-1,+1]$. 
\end{itemize}
So it is only the DEC that leads to the bound $w \leq 1$, and in this case the bound is related to causality, in that the DEC ultimately comes from asserting that the energy flux $S^a= T^{ab} V_b$, (measured by an arbitrary observer of 4-velocity $V$), should be timelike.
But since the DEC is not fundamental physics~\cite{Barcelo:twilight}, and because it is known that there are situations where due to vacuum polarization the DEC is violated in the test field limit~\cite{Visser:gvp1, Visser:gvp2, Visser:gvp3, Visser:gvp4, Visser:gvp-mg8}, it is preferable to avoid appealing to the DEC if other weaker arguments suffice.

In fact there is a non-trivial variant of the classical energy conditions that is weaker than any of the standard conditions but still strong enough for the job at hand. Let $V^a$ be any timelike 4-velocity, then $F^a = T^a{}_b \, V^b$ can be physically interpreted as the energy-momentum flux measured by an observer of 4-velocity $V$. Let us demand that this flux be causal, that is, timelike or null. This corresponds to energy flowing at speeds less than or equal to that of light, but \emph{without} any assumption that energy densities need necessarily be positive. (It is this freedom that makes the current proposal significantly weaker than the DEC.) Assuming that this ``flux energy condition'' [FEC] holds for arbitrary $V$ is equivalent to $V^a (T_a{}^c\,T_{cb}) V^b \leq 0$, which for perfect fluid stress energy tensors is equivalent to $\rho^2 \geq p^2$, that is $w \in [-1,+1]$. For more general diagonalizable stress energy tensors $T_{ab}=\hbox{diag}\{\rho,p_i\}$, this is equivalent to $\rho^2 \geq p_i^2$, which at a classical level forbids tachyons. Recall that for tachyons $\omega^2 = k^2- |\mu|^2 < k^2$, so for an isotropic tachyon gas $\rho<{1\over 3} p$. This includes the extreme case $\rho=0$ with $p\neq0$, which would correspond to a zero temperature tachyon gas.  In short, the flux energy condition enforces causal energy fluxes, and this causality-inspired constraint is sufficient to guarantee $|w|\leq 1$.

%------------------------------------------------------------------------------------------------------------------------------------------
\subsection{Speed of sound argument}
%------------------------------------------------------------------------------------------------------------------------------------------

Now assume a barotropic equation of state $\rho(p)$ with $\rho_0 = \rho(0) \geq 0$.
The speed of sound is $c_s^2 = \d p/\d\rho$, and to preserve causality we demand (at all pressures) 
\begin{equation}
{\d p\over\d\rho} \leq 1 \qquad \implies \qquad {\d\rho\over\d p} \geq 1.
\end{equation}
But then
\begin{equation}
\rho(p) = \rho_0 + \int_0^p {\d\rho\over \d \bar p} \; \d\bar p \geq \rho_0 + \int_0^p 1 \; \d\bar p. 
\end{equation}
That is
\begin{equation}
\rho(p) \geq \rho_0 + p,
\end{equation}
implying
\begin{equation}
w = {p\over\rho} \leq {p \over  \rho_0 + p} \leq 1.
\end{equation}
This is a nice physically based argument ensuring $w\leq 1$ (and hence $\langle w \rangle \leq 1$).  
Note that to keep the speed of sound real we  need $\d p/\d\rho$ to be positive, implying $w\geq 0$.
So as long as one has a barotropic equation of state with non-negative density at zero pressure, then the speed of sound argument implies $w\in [0,1]$. 

There is one exceptional point, $w=-1$, for which the speed of sound argument does not apply.  The exceptional point $w=-1$ corresponds to cosmological constant, so both $\d p$ and $\d \rho$ are by definition identically zero --- cosmological constant ``matter'' simply does not support sound waves, and no conclusions can be drawn for this exceptional point $w=-1$. 

%------------------------------------------------------------------------------------------------------------------------------------------
\subsection{Monotonicity of $w$}
%------------------------------------------------------------------------------------------------------------------------------------------

Note that by the definition of $w$ we have
\begin{equation}
{\d w\over\d p} = {1\over\rho} - {p\over\rho^2} {\d \rho\over\d p} \geq {1\over\rho} - {p\over\rho^2} = \rho \; (1- w) \geq 0.
\end{equation} 
So $w(p)$ is monotone increasing as a function of $p$. (Under the assumptions stated above: A causal barotropic equation of state $\rho(p)$ with $\rho_0 = \rho(0) \geq 0$.)

Note that under these conditions
\begin{equation}
{\d\over \d p}\left({1+w \over 1+3w }\right)   = {\d w\over \d p }  {\d\over\d w} \left({1+ w \over 1+3 w} \right)   = -  {2 \rho (1- w)\over(1+3 w )^2} < 0.
\end{equation}
So the pre-factor occurring in our entropy bound is seen to monotonically approach the ${1\over4}$ expected from black hole thermodynamics as the pressure increases.

%------------------------------------------------------------------------------------------------------------------------------------------
\subsection{High pressure limit}
%------------------------------------------------------------------------------------------------------------------------------------------

What can we say about
\begin{equation}
w_\infty = \lim_{p\to\infty} {p\over\rho(p)} \; ?
\end{equation}
Insofar as this limit makes any sense,  we can use (one variant of) the standard 'l~Hospital rule to deduce:
\begin{equation}
w_\infty = \lim_{p\to\infty} {p\over\rho(p)} =  \lim_{p\to\infty} {1\over \d\rho(p)/\d p} =  \lim_{p\to\infty} {\d p\over \d\rho} =  \lim_{p\to\infty}  c_s^2(p) = c_s^2(\infty) \leq 1.
\end{equation}
In fact, physically there are good reasons to expect
\begin{equation}
c_s^2(\infty) = 1.
\end{equation}

%------------------------------------------------------------------------------------------------------------------------------------------
\section{Positivity of the chemical potential}
%------------------------------------------------------------------------------------------------------------------------------------------
\label{appendix:chemical}

Why is $\mu_\infty N \geq 0$? If there is only one particle species a simple stability argument suffices: If the chemical potential is negative then adding more particles decreases the energy and the system is unstable~\cite{Abreu1, Abreu2, Abreu3}. As there is no energy cost to generate new particles, they will be spontaneously created. A more precise statement, however, is that once one sums over all particle species
\begin{equation}
\sum_i \mu_i N_i \geq 0.
\end{equation}
In terms of the volume number density this assertion becomes
\begin{equation}
\sum_i \mu_i n_i \geq 0.
\end{equation}
To see why this happens, consider a particle with spin degeneracy $g$. In terms of the phase-space number density we have
\begin{equation}
n_i = \int {\d^3 p\over (2\pi \hbar)^3} \; g \; \mathfrak{n}_i(p); \qquad \mathfrak{n}_i(p) = {1\over \exp\{\beta[E_i(p)-\mu_i]\}\pm 1}   
\end{equation}
\begin{itemize}
\item 
To prevent singularities, and to keep the phase-space number density $\mathfrak{n}$ positive, for bosons we must have $\mu_i \leq m_i$. (See for instance~\cite{Haber}.) For fermions there is principle no restriction.
\item
Based ultimately on CPT, for antiparticles (either fermionic or bosonic) we must have
\begin{equation}
\bar\mu_i = - \mu_i,
\end{equation}
implying (for bosons)
\begin{equation}
|\mu_i| \leq m_i.
\end{equation}
\item 
In particular, for any particle that is its own antiparticle (such as, for instance,  photons) we have
\begin{equation}
\mu_\gamma = 0.
\end{equation}
\end{itemize}
The remaining terms in the sum decompose to a sum over particle-antiparticle pairs
\begin{equation}
\sum_i \mu_i n_i = \sum_p \mu_p (n_p - n_{\bar p}).
\end{equation}
Let us focus on the phase space number density $\mathfrak{n}$. Then consider the combination
\begin{eqnarray}
\mu_p (\mathfrak{n}_p - \mathfrak{n}_{\bar p}) &=& \mu ( \mathfrak{n} - \bar{\mathfrak{n}}) 
\\
&=& \mu \left[ {1\over \exp\{\beta[E-\mu]\}\pm 1}- {1\over \exp\{\beta[E+\mu]\}\pm 1} \right]
\\
&=& \mu { \exp\{\beta[E+\mu]\} - \exp\{\beta[E-\mu]\} \over [ \exp\{\beta[E-\mu]\}\pm 1][ \exp\{\beta[E+\mu]\}\pm 1] }
\\
&=& \mu {2 \sinh \{\beta\mu\} \exp\{\beta E\}\over \exp\{2\beta E\} \pm 2\exp\{\beta E\}\cosh\{\beta\mu\} + 1}
\\
&=&  {\mu \sinh \{\beta\mu\} \over \cosh\{\beta E\} \pm \cosh\{\beta\mu\} } 
\\
&\geq& 0.
\end{eqnarray}
We emphasise that this final inequality holds \emph{regardless of the sign} of $\mu$. The denominator is always positive for both fermions and bosons (in view of the fact that for bosons $|\mu|\leq m$.)
Normally there are more particles than antiparticles and so the chemical potential of the particles is positive. But even in ``reversed'' situations (more antiparticles than particles) we will still have $\mu ( \mathfrak{n} - \bar{\mathfrak{n}}) \geq 0$. 
Integrating over momentum space, for each pair of particle antiparticle species we have
\begin{equation}
\mu ( n - \bar n) \geq 0.
\end{equation}
Summing over all particle antiparticle pairs of species we see
\begin{equation}
\sum_i \mu_i n_i = \sum_p \mu_p (n_p - n_{\bar p}) \geq 0.
\end{equation}
This implies
\begin{equation}
\sum_i \mu_i N_i  \geq 0.
\end{equation}
We usually summarize this by saying $\mu_\infty N\geq 0$, though a more precise statement would be that the (net) average chemical potential is always positive:
\begin{equation}
\langle \mu\rangle = {\sum_i \mu_i N_i \over \sum_i  N_i } \geq 0.
\end{equation}

%-----------------------------------------------
\section*{Acknowledgements}
%-----------------------------------------------

Gabriel Abreu and Matt Visser were supported by the Marsden Fund administered by the Royal Society of New Zealand.  Gabriel Abreu was also supported by a Victoria University Postgraduate Scholarship. Carlos Barcel\'o was supported by Spanish MICINN through the project FIS2008-06078-C03-01 and by the Junta de Andaluc{\'\i}a through the project FQM219.

\enlargethispage{30pt}
%-------------------------------

%-------------------------------
%----------------------------------------------------
\end{document}